\RequirePackage{fixltx2e}
\documentclass[aps,jmp,prl,amsmath,amssymb,twocolumn,
english,floatfix]{revtex4-1}
\usepackage[colorlinks, linkcolor=blue,
anchorcolor=blue, citecolor=blue]{hyperref}
\usepackage{graphicx}
\usepackage{dcolumn}
\usepackage{bm}
\usepackage{epstopdf}
\usepackage{multirow}
\usepackage{amssymb}
\usepackage{xcolor} 
\usepackage[normalem]{ulem}

\begin{document}
\bibliographystyle{physrev}
\title{Native surface oxide turns alloyed silicon membranes into nanophononic metamaterials with ultra-low thermal conductivity}
\author{Shiyun Xiong$^{1,2}$}
\email{xiongshiyun216@163.com}
\author{Daniele Selli$^{2}$}
\author{Sanghamitra Neogi$^{3}$}
\author{Davide Donadio$^{4}$}
\email{ddonadio@ucdavis.edu}
\affiliation{1 Functional Nano and Soft Materials Laboratory (FUNSOM) and Collaborative Innovation Center of Suzhou Nano Science and Technology, Soochow University, Suzhou, Jiangsu 215123 , P.R. China\\
2 Max Planck Institute for Polymer Research,  Ackermannweg 10, 55218 Mainz, Germany\\
3 Department of Aerospace Engineering Sciences, University of Colorado Boulder, Boulder, Colorado 80309, USA\\
4 Department of Chemistry, University of California Davis, One Shields Ave. Davis, 95616, CA
}

\begin{abstract}
A detailed understanding of the relation between microscopic structure and phonon propagation at the nanoscale is essential to design materials with desired phononic and thermal properties.
Here we uncover a new mechanism of phonon interaction in surface oxidized membranes, i.e., native oxide layers interact with phonons in ultra-thin silicon membranes through local resonances. The local resonances reduce the low frequency phonon group velocities and shorten their mean free path. This effect opens up a new strategy for ultralow thermal conductivity design as it complements the scattering mechanism which scatters higher frequency modes effectively. 
The combination of native oxide layer and alloying with germanium in concentration as small as 5\% reduces the thermal conductivity of silicon membranes to 100 time lower than the bulk. In addition, the resonance mechanism produced by native oxide surface layers is particularly effective for thermal condutivity reduction even at very low temperatures, at which only low frequency modes are populated.
\end{abstract}
\maketitle

Controlling terahertz vibrations and heat transport in nanostructures has 
a broad impact on several applications, such as thermal management in micro- and nano-electronics, renewable energies harvesting, sensing, biomedical imaging and information and communication technologies~\citep{Pop:2010dv,RN133,ZhuNL2014,CottrillAEM2015,LiPRL2004,LiRMP2012,ZhuAS2016,Volz:2016ky}.
Significant efforts have been made to understand and engineer heat transport in nanoscale silicon due to its natural abundance and technological relevance \citep{Lee2012,ChihweiNN,Boukai2008,AllonNat2008}. In the past decade researchers explored strategies to obtain silicon based materials with low thermal conductivity (TC) and unaltered electronic transport coefficients, so to achieve high thermoelectric figure of merit and enable silicon-based thermoelectric technology\citep{Boukai2008,AllonNat2008,Joshi:2008dd,Tang:2010kv,Aksamija2010,Boor2012,YuBoNL2012,XiongSMLL2014}.

From the earlier studies it was recognized that low-dimensional silicon nanostructures, such as nanowires, thin films and nano membranes feature a largely reduced TC, up to 50 times lower than that of bulk at room temperature. TC reduction becomes more prominent with the reduction of the characteristic dimension of the nanostructures~\cite{DeyuAPL03,Ju:1999uy,Liu:2011dx,ChavezAngel:2014be}.
Theory and experiments consistently show that surface disorder and the presence of disordered material at surfaces play a major role in determining the TC of nanostructures \cite{AllonNat2008,donadioPRL09,He:2012dn,Neogi2015}. However, a comprehensive understanding of the physical mechanisms underlying so large TC reduction is lacking. The effect of surface roughness and surface disorder on phonons has been so far interpreted in terms of phonon scattering \cite{Martin:2009hl,Chen:2008ig,Kazan:2010tz,Johnson:2013ic,Maznev:2015di}, but  scattering would not account for mean free path reduction of long-wavelength low-frequency modes.
Recent theoretical work  demonstrated that surface nanostructures, such as nanopillars at the surface of thin films or nanowires, can efficiently reduce TC through resonances, a mechanism that is intrinsically different from scattering~\cite{DavisPRL2014,XiongPRL2016}. Surface resonances alter directly phonon dispersion relations by hybridizing with propagating modes in the nanostructures, thus hampering their group velocity. 
 

In this Communication we unravel the effect of  native oxide surface layers on thermal transport in ultra-thin silicon membrane models that closely resemble experiments, 
by atomistic molecular dynamics simulations. We show that the observed low TC in these systems, and plausibly in other oxide coated silicon nanostructures, is predominantly due to resonances analogous to those occurring in nanophononic metamaterials. It is worth stressing that  surface oxide layers in low dimensional Si materials grow spontaneously at atmospheric conditions, and do not require any specific processing.
Our simulations ascertain the occurrence of low frequency resonant modes that hybridize with the acoustic branches ($\omega\lesssim4$ THz)  of the membrane, effectively suppressing their mean free path (MFP). This discovery opens up the possibility to further optimize the TC of ultra-thin silicon membranes by combining resonances with mass scattering, which affects phonons with higher frequency ($\omega\gtrsim4$ THz). We show that alloying the crystalline core of ultra-thin membranes with a small percentage of substitutional germanium atoms brings forth ultra-low TC in silicon membranes with technologically viable thickness~\citep{ShcheAPL2013}.


All TCs are calculated with equilibrium molecular dynamics (EMD) simulations at 300 K using LAMMPS \citep{LAMMPS} with interatomic interactions described by the widely used Tersoff potential \citep{TersoffSiD,TersoffSiO,TersoffGeO}. The equations of motion are integrated with the velocity Verlet algorithm with a time step of 0.8 fs for the membranes without surface oxide. Due to the stiffness of the Si-O bond and the light weight of oxygen, a timestep of 0.12 fs is used to guarantee energy conservation in systems with surface oxide. The length-dependent phonon transmission functions that provide phonon MFPs, are obtained from non-equilibrium molecular dynamics (NEMD) simulations~\cite{Jund:1999tb}.  Surface oxidized samples are about 1 nm thicker than the  corresponding pristine Si membranes due to expansion of the oxide layer. To simplify the notation and also facilitate comparisons, we use the original Si membrane thickness to denote the thickness of surface oxidized membranes (real thicknesses are used for all calculations). Simulation details and procedures to generate models with native oxide layers are reported in the supplementary materials \cite{Suppl}.

 Fig.~\ref{TC} shows the TC of the pristine and the surface oxidized Si membranes, together with the TC of Si$_{1-x}$Ge$_x$ membranes with Ge concentration $x$ ranging from 0.05 to 0.2 \footnote{All the TCs reported here are the TCs of the whole sample, including the surface oxide if involved.}. The TC of crystalline Si membranes with (2x1) surface reconstruction and with native oxide surface layers were calculated in a former study by EMD~\cite{Neogi2015}. It was revealed that the simulated TCs of Si membranes with ideal surfaces is larger than the experimental results \citep{ChavezAngel:2014be,Cahill2003}. In turn the TC of models with native oxide surface layers is in excellent agreement with experiments~\cite{Neogi2015}: $\sim1$ nm oxide layers abate the TC by about 4, 12 and 17 times for 10.9 nm, 5.4 nm, and 3.2 nm thick membranes , respectively, compared to the corresponding ones with pristine surface reconstruction. The obtained TC of the surface oxidized membranes is much smaller than that estimated by effective medium theory \citep{EMT-Nan} as shown in table S1.
\begin{figure}[!htb]
\centering
\includegraphics[scale=0.53]{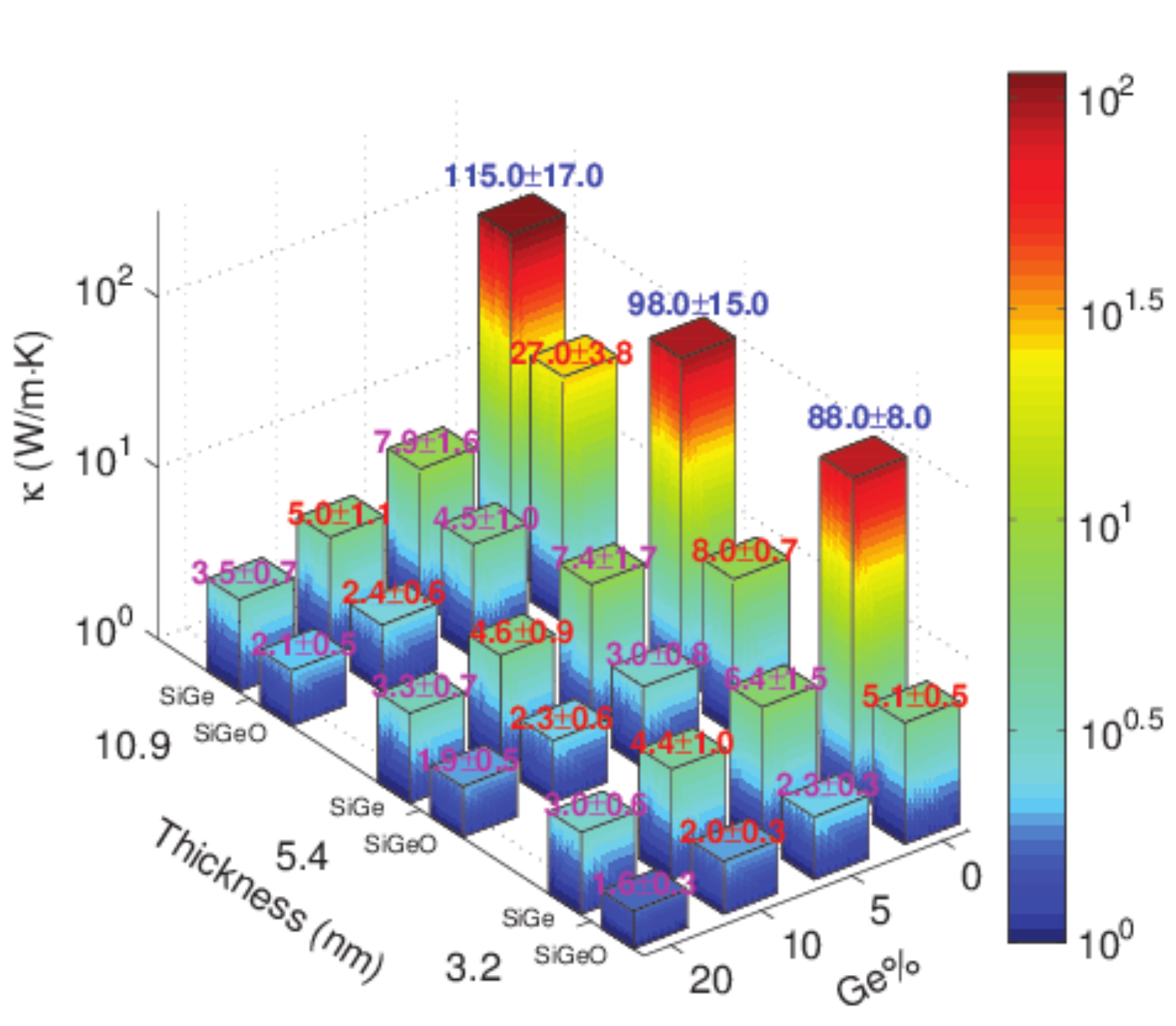}
\caption{Thermal conductivities of silicon membranes as functions of thickness and Ge concentration at 300 K. Si$_{1-x}$Ge$_x$ membranes with ordered $2\times 1$ surface reconstruction (labeled as SiGe) are compared to those covered by native oxide layers (labeled as SiGeO). The error bars are calculated as the standard deviation.}
\label{TC}
\end{figure}

Ge alloying, even at fairly low concentration, promotes a dramatic reduction of TC, not only in the membranes with pristine surfaces, which feature high TC, but also in the ones covered with native oxide. The effect of alloying on the TC of ultra-thin membranes is more prominent than in bulk \citep{SiGekappa} and it is analogous to the effect observed in nanowires~\cite{Zhang:2010ez}, where TC keeps decreasing even at Ge concentrations for which the bulk alloy TC has saturated. 
When alloying is combined to surface oxidation, extremely low values of TC are observed: the TC of the 10.9 nm thick membrane drops to 4.5 Wm$^{-1}$K$^{-1}$ with only 5$\%$ of Ge and to 2.1 Wm$^{-1}$K$^{-1}$ with a Ge concentration of 20$\%$. The lowest TC (1.6 W/mK)  achieved in the 3.2 nm thick surface oxidized Si$_{0.8}$Ge$_{0.2}$ is much lower than that of bulk amorphous Si \citep{HeypAPL2011}, thus achieving effectively the phonon-glass paradigm in a crystalline material \cite{Mahan:1996te}.  

According to former first-principles calculations, Ge alloying strongly reduces the mean free paths of medium and high frequency phonons~\cite{Garg:2011hi}. Similarly, based on the scattering picture \cite{Maznev:2015di}, surfaces would affect phonons in the same range of frequencies, i.e. modes with wavelengths comparable to the characteristic length of surface features, namely the interfacial and surface roughness, which are in the order of few nm. 
While further large TC reduction ($\sim$ a factor of 2) produced by surface oxide layers in alloyed membranes suggests that native oxide should hinder the low frequency phonon transport, which cannot be accounted by surface scattering. 
As a result, a different mechanism to describe the role of surface oxide on TC reduction need to be clarified. 
Hereafter we offer a detailed explanation of such mechanism, and we elucidate the concerted effects of mass scattering and surface oxidation on phononic transport.
We first resolve the frequency ranges that are affected by either native oxide layers or alloying. To this aim we calculate the  phonon MFP from length dependent phonon transmission function $\mathcal{T}(\omega)$. For the transmission functions, we exploit an NEMD-based approach developed recently by S\"a\"askilahti {\it et al.} \citep{KimmoPRB2014,KimmoPRB2015} to take into account anharmonic effects at a reasonable computational cost. MFPs $\Lambda(\omega)$  at each frequency are fitted according to the transmission functions at each length, which are averaged over at least 10 data sets (Supplementary material \cite{Suppl}):

\begin{figure}[!htb]
\centering
\includegraphics[scale=0.4]{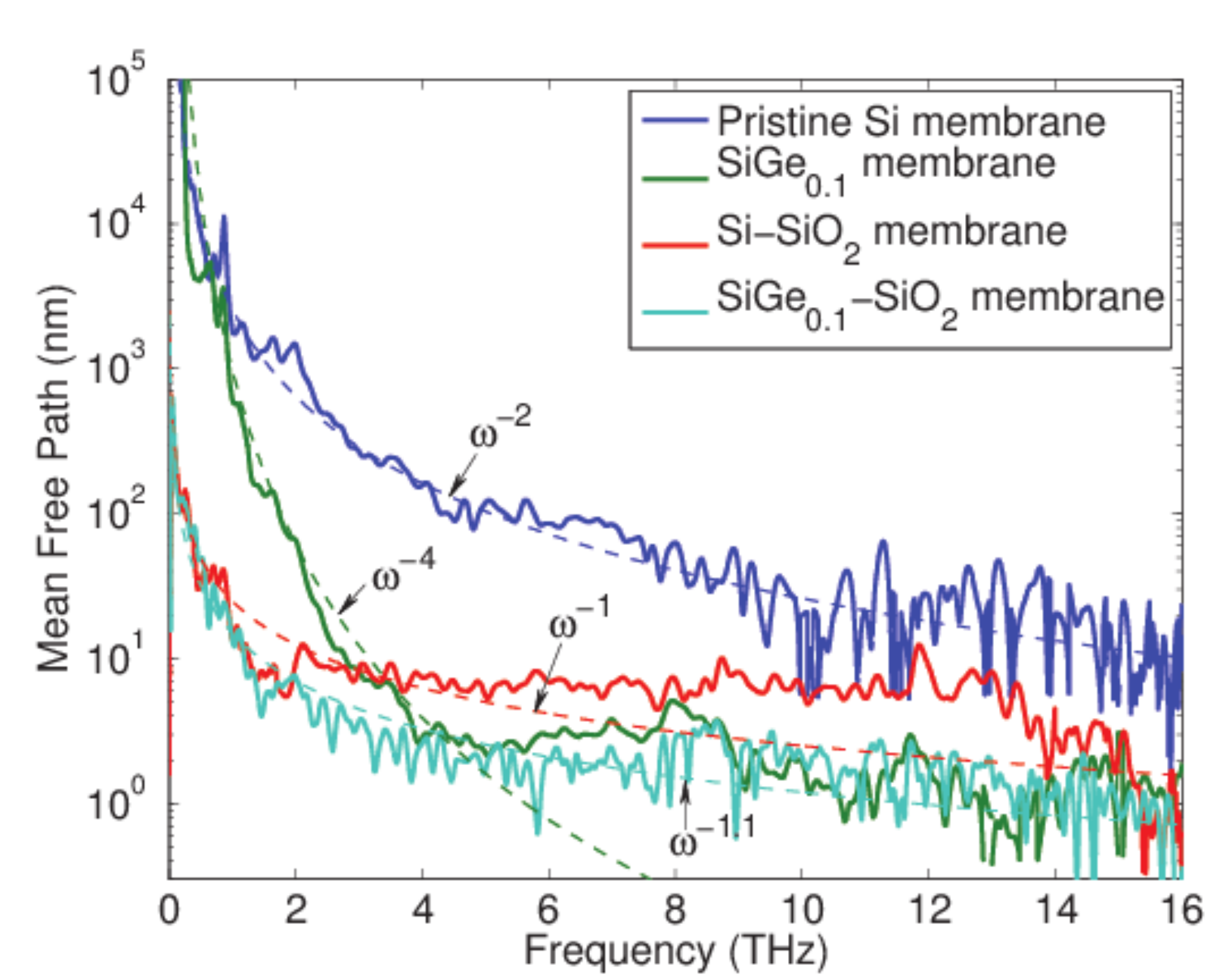}
\caption{Phonon mean free paths evaluated from the length dependent transmission functions for the following membranes of 3.2 nm in thickness: pristine Si membrane, SiGe$_{0.1}$ membrane, amorphous SiO$_2$ covered Si membrane, and amorphous SiO$_2$ covered SiGe$_{0.1}$ membrane. }
\label{MFP}
\end{figure}
Fig. \ref{MFP} shows the MFPs obtained from the transmission functions computed for 3.2 nm thick Si and Si$_{0.9}$Ge$_{0.1}$ membranes. 
For the case of pristine Si membrane, the MFP is relatively large in the whole frequency range and it is proportional to $\omega^{-2}$, as a consequence of phonon-phonon scattering~\cite{Callaway:1959uo,Holland:1963uo}. 
The fitted behavior reveals that boundaries almost do not scatter phonons and are highly specular, which is consistent with the observed large and weakly thickness dependent TCs in pristine Si membranes. 
With 10\% random Ge alloyed into the Si membrane (Si$_{0.9}$Ge$_{0.1}$  membrane), the MFPs are suppressed by about two orders of magnitude for $\omega$ \textgreater{} 2 THz. The MFPs above 4 THz are even shorter than 3 nm, thus those phonons contribute little to the total TC. In contrast, MFPs below 1 THz are not affected by alloying and those phonons can travel longer than micrometers without being scattered, which is in agreement with experiments showing several $\mu$m long MFP in SiGe nanowires \citep{ChihweiNN}.
 Since phonons beyond 4 THz in Si$_{0.9}$Ge$_{0.1}$ membrane almost do not contribute to the total TC, we fitted the MFP below 4 THz with $\omega^{-4}$, which is in accordance with kinetic theory \cite{Holland:1963uo}.
In contrast to alloying, surface oxidation also suppresses the MFP of low frequency phonons below 4 THz by more than 2 orders of magnitude. At higher frequency, the effect of native oxide layer on MFP is instead weaker than that of alloying. 
With surface oxide, $\Lambda(\omega)$ is proportional to $1/\omega$ at low frequency. The large suppression of low frequency phonon MFP makes the resonant structures very efficient at reducing TC at extremely low temperatures \citep{Reson1976,MaasiltaNC2014}, at which only low frequency modes that are insensitive to mass scattering are populated. The different frequency range response observed in membranes upon surface oxidation and alloying makes them complement each other. This feature enables the suppression of MFPs over the whole phonon spectrum when alloying and native surface oxidation are combined, as evidenced by the MFP of surface oxidized Si$_{0.9}$Ge$_{0.1}$ membrane. The suppressed MFP in the whole frequency range finally leads to the observed low TC in surface oxidized SiGe membranes. The ability of blocking both low and high frequency phonons by combining surface oxidation and alloying sets a valuable strategy for the design of ultra-low TC nanostructures.
%

As scattering relies upon the particle-like nature of phonons, it affects only modes with wavelength comparable to the characteristic length of the scatterers. Our  calculated MFPs show that native oxide layers mainly affect phonons at low frequencies, with wavelength much longer than the characteristic interfacial and surface roughness, which is $\sim 1$ nm. Those phonons typically have MFPs ranging from hundreds of nm to several $\mu$m, thus contributing significantly to the total TC. In addition, a portion of the phononic spectrum of these systems still exhibits MFP much larger than the so called Casimir limit--corresponding to the thickness of the membranes \cite{Casimir:1938tf}. 
In general, phonons with frequency below 1 THz have wavelengths larger than several tens of nm and cannot be treated as particles, as assumed in scattering theories. As a result, the commonly employed scattering mechanism cannot explain the frequency dependent MFP reduction induced by oxidized surfaces and new mechanisms have to be explored.
Recent works suggested that resonant structures may provide a viable source of TC reduction in nanostructures \cite{DavisPRL2014,XiongPRL2016}. Different from scattering, the resonant mechanism relies on the wave-like nature of phonons. Localized resonant modes have a great effect on TC by interacting with the propagating phonons in the thin films and nanowires, modifying their dispersion relations and reducing their group velocity, especially at low frequency. Because of the lack of periodicity of the amorphous SiO$_2$ surface, the produced resonances are distributed randomly. A similar effect was also proposed for core-shell Si/Ge nanowires, in which resonances occur due to the mismatch of group velocities in the core and shell regions \citep{JChenNL2012,JChenJCP2011}.

To prove that the MFP reduction at low frequency is indeed due to random resonant modes, we obtained the effective phonon dispersion relations of pristine and  surface oxidized Si membranes, computing the dynamical structure factor (DSF) in an EMD simulation \citep{strufac}:
\begin{equation}
S\left(\vec{q},\omega\right) = \frac{1}{2\pi}\left| \int_0^{T_0} \sum_1^N \exp \left[-i\vec{q}\cdot \vec{r}_j(t)\right] \exp\left(i\omega t\right)dt \right|
\label{DSF}
\end{equation}
where $\vec{q}$ is the phonon wave vector. $\vec{r}_j(t)$  refers to the time (\emph{t}) dependent position of atom \emph{j}. $T_0$ and \emph{N} denote the total simulation time and the number of atoms, respectively.
\begin{figure}[!htb]
\centering
\includegraphics[scale=0.2]{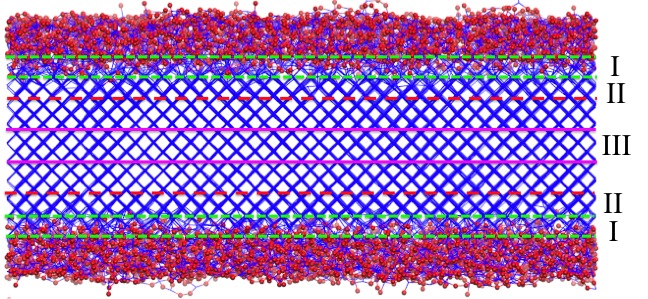}
\includegraphics[scale=0.4]{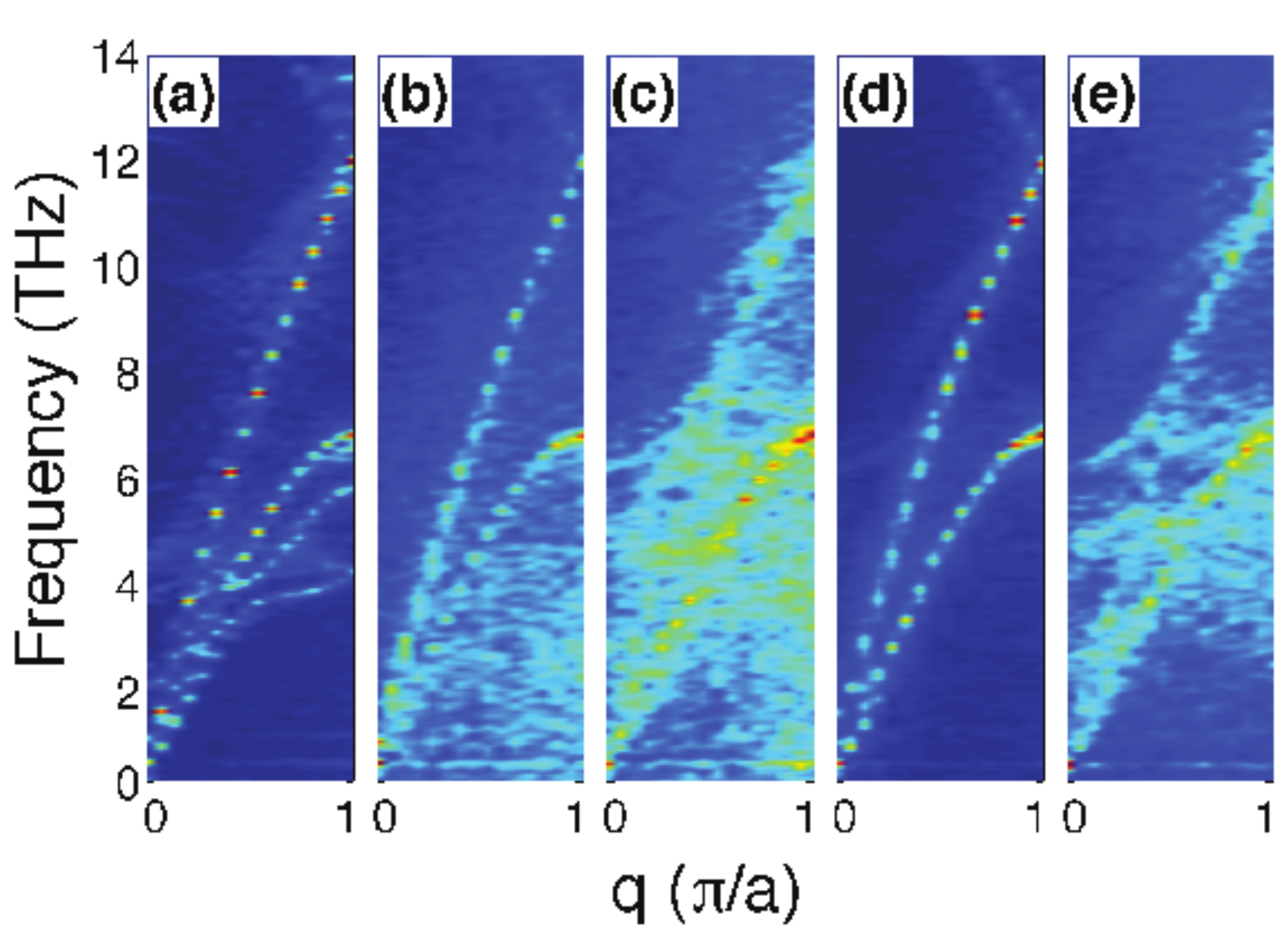}
\caption{Schematic of a surface oxidized Si membrane and the definition of regions for DSF calculations (top). Dispersion of acoustic phonons as calculated from DSF (down): (a) pristine Si membrane; surface oxidized Si membrane calculated from (b) the whole sample, (c) region II (the region next to the interface), and (d) region III (middle Si region away from interfaces); (e) partial DSF of region II with 10 holes introduced in the interface region I.}
\label{disp}
\end{figure}
Fig. \ref{disp} shows the phonon dispersion curves calculated with the DSF (Eq. \ref{DSF}) of different cases. For the pristine Si membrane (Fig. \ref{disp}(a)), the dispersion show clearly the longitudinal acoustic (LA) and transverse acoustic (TA) branches. We note that in additional to the normal LA and TA modes, surface branches, typically occurring in membranes, are also visible~\cite{Neogi2015}. 
In the system with native oxide at the surfaces, the TA and LA modes show similar behavior as the pristine membrane, but, besides the TA and LA modes, numerous dispersion-less modes appear at random frequencies, especially below 8 THz. 
A similar observation was made for SiO$_2$ coated SiNWs \citep{strufac}. 
Such dispersionless modes are resonant modes due to the vibrations of the atoms in the amorphous SiO$_2$ layers. Random resonances produced by the amorphous SiO$_2$ layers can penetrate into the crystalline Si region and interact with the LA and TA modes, thus reducing the group velocity of propagating modes. The penetration of the resonant modes can be examined by comparing the partial DSF of the regions next to the interfaces (region II, Fig. \ref{disp}(c)) to the middle Si region (region III, Fig. \ref{disp}(d)). 
Random resonant modes are mostly localized in the first few layers near the interface (Region II in Fig. \ref{disp}). The hybridization strength between resonant and propagating modes is proportional to the coupling strength between the oxide layers and the Si core. To probe the effect of varying such coupling strength, we introduce 10 pores of 2.5 {\AA} in diameter in the interface region (region I). The pores are equally spaced and are aligned across one periodic edge of the simulation box. After introducing the pores, we observe that the number of  dispersion-less modes in region II (Fig. \ref{disp}(e)) is reduced, compared to the system without pores (Fig. \ref{disp}(c)), revealing a weaker interaction between resonating and propagating modes.
The ability of modifing phonon dispersion through local resonances tunes native surface oxidized membranes to be a nanophononic material. This new type of nanophononics does not require any periodic structure and it is different from the conventional periodic nanophononics, thus facilitating the fabrication process.

\begin{figure}[!htb]
\centering
\includegraphics[scale=0.25]{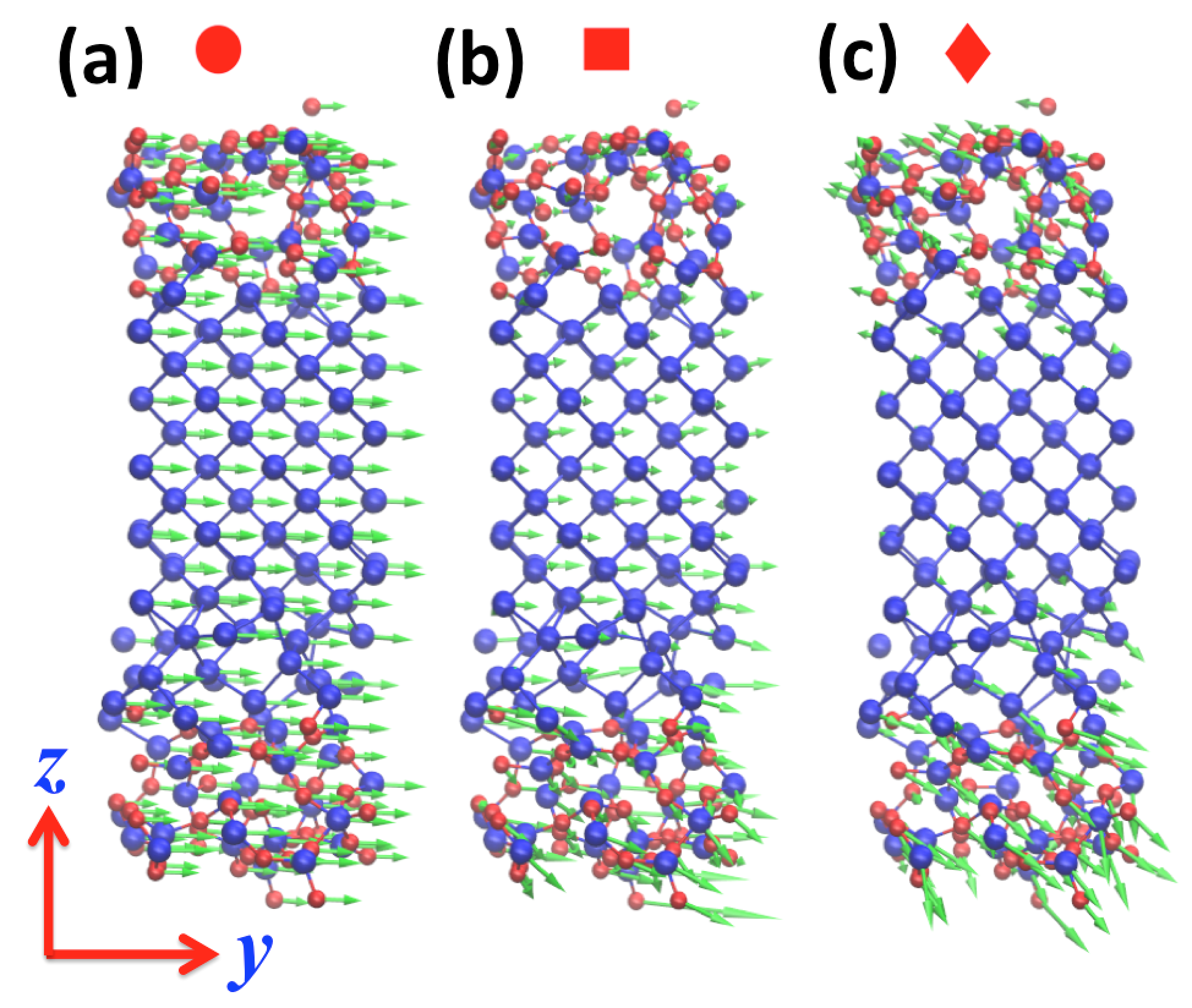}
\caption{Visualization of three selected modes on the same branch by VMD software \citep{VMD}. (a) a transverse acoustic mode at wavevector $q=$0.2; (b) a hybridized mode at $q=$0.6 resulting from the hybridization between a transverse acoustic mode and a surface resonant mode; (c) $q=$0.95, a resonant mode localized in the native oxide layers. The wavevector is perpendicular to the y-z plane. }
\label{vector}
\end{figure}

To directly illustrate the resonant modes and their hybridization with the propagating phonons, we have calculated the vibrational eigenvectors by lattice dynamics \citep{BornH} for the surface oxidized Si membrane of 2$\times$2 unit cell in periodic directions and of 3.2 nm in thickness. Note that the cell in periodic directions is too small to capture the ture amorphous SiO$_2$, while it can still produce some resonances and show the hybridization effect. Fig. \ref{vector} visualizes three phonon modes of the same branch at $q =$ 0.2, 0.6 and 0.95, which demonstrates the hybridization effect bewteen the TA mode and a resonant mode. The notations (circle, square and diamond) correspond to the red notations of the same shape in Fig. S2 \citep{Suppl}).

As demonstrated in Fig. \ref{vector}, at $q =$ 0.2, all atoms vibrate along the transverse direction with the same magnitude (Fig. \ref{vector}(a)) as in a pure TA mode. Since the vibration of all atoms are in phase, such a mode can propagate with a large group velocity. 
For the same branch at $q =$ 0.95, only the atoms belonging to  the surface oxide layers participate to the vibration, indicating that the mode is a localized resonance (\ref{vector}(c)). Such a mode almost does not propagate and is able to hybridize with the TA mode, which is illustrated by the mode at  $q =$ 0.6 (Fig. \ref{vector}(b)). For this mode, the vibrational amplitude increases gradually from the top surface to the bottom surface,  due to the fact that the atom vibrations of the resonant mode on the top and bottom surfaces are in opposite directions; the hybridization between the resonance and LA mode finally resulted in large vibrational mismatch among different atoms, which reduces the group velocity of the original propagating mode.

In conclusion, through a detailed study of heat transport in surface oxidized and alloyed membranes and of the corresponding mechanisms, we conclude that instead of scattering the high frequency phonons, the main effect of native  oxide surface layers on TC reduction is to limit the propagation of low frequency acoustic modes in ultra-thin silicon membranes due to hybridization with localized resonances. The mechanism can be effectively combined with mass scattering, which strongly hampers the MFP of higher frequency phonons. Thus the two mechanisms can be combined to design nanostructures with extremely low TC, and make it possible to produce crystalline structures with TC below the corresponding amorphous value. TC from 50 to 100 times lower than that of bulk silicon at room temperature are predicted for Ge concentration as low as 5$\%$, thus enabling efficient thermoelectric energy conversion. Assuming a low impact of Ge on charge transport \cite{Mangold}, it would be possible to obtain ZT \textgreater 1 in $\sim$7 nm thick membranes. 


We are indebted to Kimmo S\"a\"askilahti for useful discussions and the assistance of transmission calculations. Yuriy A. Kosevich and Mahmoud I. Hussein are acknowledged for fruitful discussions. S.X. acknowledges the scholarship from the Alexander von Humboldt foundation. This work is supported by Jiangsu provincial natural science funding project BK20160308. Simulations are performed on the cluster HYDRA from Rechenzentrum Garching of the Max Planck Society (MPG).

\bibliographystyle{apsrev4-1}

%

\end{document}